\documentclass[10pt,conference]{IEEEtran}
 \usepackage[]{todonotes} 
\usepackage[ruled, vlined, linesnumbered, noend]{algorithm2e}
\usepackage[super]{nth}
\usepackage{mathtools}
\usepackage{xcolor}
\usepackage{booktabs}
\usepackage{amsfonts}
\usepackage{subcaption}
\usepackage{flushend}

\usepackage{xspace}

\def\@copyrightspace{\relax}

\definecolor{amber}{rgb}{1.0, 0.75, 0.0}
 
\newcommand{\ballnumber}[1]{\tikz[baseline=(myanchor.base)] \node[circle,fill=.,inner sep=1.3pt] (myanchor) {\color{white}\bfseries\footnotesize #1};}
\newcommand{\yellowballnumber}[1]{\tikz[baseline=(myanchor.base)] \node[circle,fill=amber,inner sep=1.3pt] (myanchor) {\color{black}\bfseries\footnotesize #1};}

\newcommand{\namex}{NetNN\xspace}

\begin{document}
\title{\namex: Neural Intrusion Detection System in Programmable Networks}
\author{\IEEEauthorblockN
	{Kamran Razavi$^{\dagger}$$^{\S}$,~Shayan Davari Fard$^{\dagger}$,~George Karlos$^{\ast}$,~Vinod Nigade$^{\ast}$,~Max M\"uhlh\"auser$^{\dagger}$,~Lin Wang$^{\S}$\\}
	\vspace{0.2cm}
	\IEEEauthorblockA{$^{\dagger}$Technical University of Darmstadt, $^{\ast}$Vrije Universiteit Amsterdam, $^\S$Paderborn University}
}

\maketitle
\begin{abstract}



The rise of deep learning has led to various successful attempts to apply deep neural networks (DNNs) for important networking tasks such as intrusion detection. Yet, running DNNs in the network control plane, as typically done in existing proposals, suffers from high latency that impedes the practicality of such approaches. 
This paper introduces \namex, a novel DNN-based intrusion detection system that runs completely in the network data plane to achieve low latency. \namex adopts raw packet information as input, avoiding complicated feature engineering. \namex mimics the DNN dataflow execution by mapping DNN parts to a network of programmable switches, executing partial DNN computations on individual switches, and generating packets carrying intermediate execution results between these switches. We implement \namex in P4 and demonstrate the feasibility of such an approach. Experimental results show that \namex can improve the intrusion detection accuracy to 99\% while meeting the real-time requirement.

\end{abstract}

\section{Introduction}

In network security, Intrusion Detection Systems (IDSes) play an important role in identifying and responding to anomalous behaviors in network traffic.
Traditional IDSes commonly employ statistical methods which rely on a prior knowledge of known attack patterns~\cite{zhang2020poseidon, liu2021jaqen}. However, these methods are limited in detecting unknown attacks, necessitating a more advanced and adaptive approach. Recently, machine learning (ML) approaches, such as Decision Trees (DTs) and Random Forests (RFs), have made their way into networking tasks such as intrusion detection~\cite{barradas2021flowlens, zheng2022iisy, zhou2023efficient, akem2023flowrest}. While using DTs and RFs reduces the required expertise in the field and provides insights about the decisions similar to rule-based approaches, it still requires a critical step called feature engineering---selecting the best features among many to build the ML model. This step is known to suffer from missing or sensitive features, and is prone to overfitting. 

Deep learning has gained tremendous attention due to its successes in various applications such as speech recognition, computer vision, and natural language processing~\cite{2020-osdi-serving, razavi2022fa2}. Deep neural networks (DNNs) can automatically extract features from large and complex data sets, and learn nonlinear relationships between input and output variables, making them a well-suited approach for complex decision-making applications such as detecting intrusions in network traffic. As a result, several successful attempts have been made to build IDSes based on DNNs, with significant detection accuracy improvement~\cite{xing2020netwarden, doriguzzi2024introducing, 2022-asplos-taurus}. However, these attempts implement the DNN-based intrusion detection models in the network control plane, incurring high latency and thus failing to meet stringent timing requirements.

Over the last decade, programmable network devices (e.g., programmable switches and SmartNICs) have been increasingly adopted in modern networks, allowing to implement customized network functions in the network data plane. This has not only accelerated network innovations~\cite{2019-hotnets-classification, 2022-nsdi-n3ic, 2022-asplos-taurus, razavi2022distributed}, but also inspired new directions such as in-network computing where computations (e.g., aggregation, caching, and coordination) are offloaded to the network data plane for low-latency high-throughput processing~\cite{2017-sosp-netcache,2018-nsdi-netchain,2021-nsdi-switchml}. 

Given the ever-increasing need for real-time network security and rapid advancements in programmable networks, it is imperative to explore the potential of leveraging programmable networks for efficient intrusion detection using DNNs. This raises an important question: Can we harness the capabilities of programmable networks, such as the packet processing speed and nanosecond-level latency offered by modern programmable switches, to effectively perform DNN-based intrusion detection in a timely and efficient manner? DNN serving would achieve an unprecedented level of performance if the DNN is executed entirely in the network data plane. Since programmable switches are natively responsible for moving the DNN input data, with this design, the inference can be performed ``on the fly,'' eliminating the need for external accelerators. However, to achieve inference at the line rate, it is essential to have pure data-plane DNN implementations which expose unique challenges due to complex DNN computations and limited hardware capabilities.

We present \namex{}, the first intelligent data-plane system that enables the execution of DNNs using a set of programmable switches for intrusion detection. To this end, \namex{} \ballnumber{1} maps the DNN neurons and associated weights to a set of programmable switches, \ballnumber{2} mimics the execution workflow of the DNN and routes the DNN execution results with network packets, and \ballnumber{3} issues instructions to individual switches, wherein each switch is tasked with executing the computations dictated by the neurons specifically assigned to it. Overall, this paper makes the following contributions. After presenting the motivation for IDSes in network data planes and identifying the challenges (Section~\ref{sec:motivation}), we 
\begin{itemize}
    \item present the design of \namex{}, including its overall architecture and components (Section~\ref{sec:design});
    \item introduce a novel network design for DNN-based IDSes that not only advances state-of-the-art intrusion detection accuracy but also enables deep learning inference execution completely in the data plane (Section~\ref{sec:mapper});
    \item implement a system prototype for \namex{}\footnote{NetNN open source code: https://github.com/shynfard/netnn}. using the P4 language, and evaluate it using a Covert Channel dataset (Section~\ref{sec:evaluation}). 
    Overall, \namex{} increases the accuracy of intrusion detection systems on programmable switches to $83\%$ without the need for feature engineering and $99\%$ by adding the hardware timestamp when packets arrive (packets' inter-arrival time) and the majority voting of the same flow's packets classification.
\end{itemize}

\section{Motivation} \label{sec:motivation}
\subsection{Feature Extraction for Intrusion Detection} \label{sec:feature_extraction}

To understand the importance of features engineering in traditional approaches for intrusion detection and the advantages of deep learning based approaches, we compare the accuracy achieved by the state-of-the-art decision-tree-based approach NetBeacon~\cite{zhou2023efficient}, an eXtreme Gradient Boosting (XGBoost)~\cite{chen2016xgboost} decision trees model with randomly selected features, and \namex, for both packet and flow classification on a convert channel detection dataset~\cite{barradas2021flowlens}.
The covert channel detection dataset consists of 1000 flows encoded by two censorship resistance tools: Facet~\cite{li2014facet} and DeltaShaper~\cite{barradas2017deltashaper}.

\noindent\textbf{Packet classification.} For \namex{}, we use the first 68 (maximum of UDP and IP header size) raw bytes of the packets (from the dataset~\cite{barradas2021flowlens}) and train a neural network with three layers (a 1D convolutional layer with 32 filters, a dense layer with 50 neurons, and another dense layer with 100 neurons) similar to~\cite{marin2019deepsec}. For NetBeacon, we train an XGBoost model using the packet's length and inter-arrival time. For the random selection approach, we select a random subset of the same features of NetBeacon and train an XGBoost model. With only the packet features (length and inter-arrival time), NetBeacon reaches the accuracy of $58\%$ using the XGBoost classifier. The accuracy drops to $51\%$ when selecting a random set ($10\%$) of lengths and inter-arrival times. On the other hand, \namex{} achieves an accuracy of $83\%$ by just using the first 68 bytes of raw representation of the packets.

\begin{figure}[!t]
    \centering
    \vspace*{0.03in}
    \includegraphics[width=0.45\textwidth]{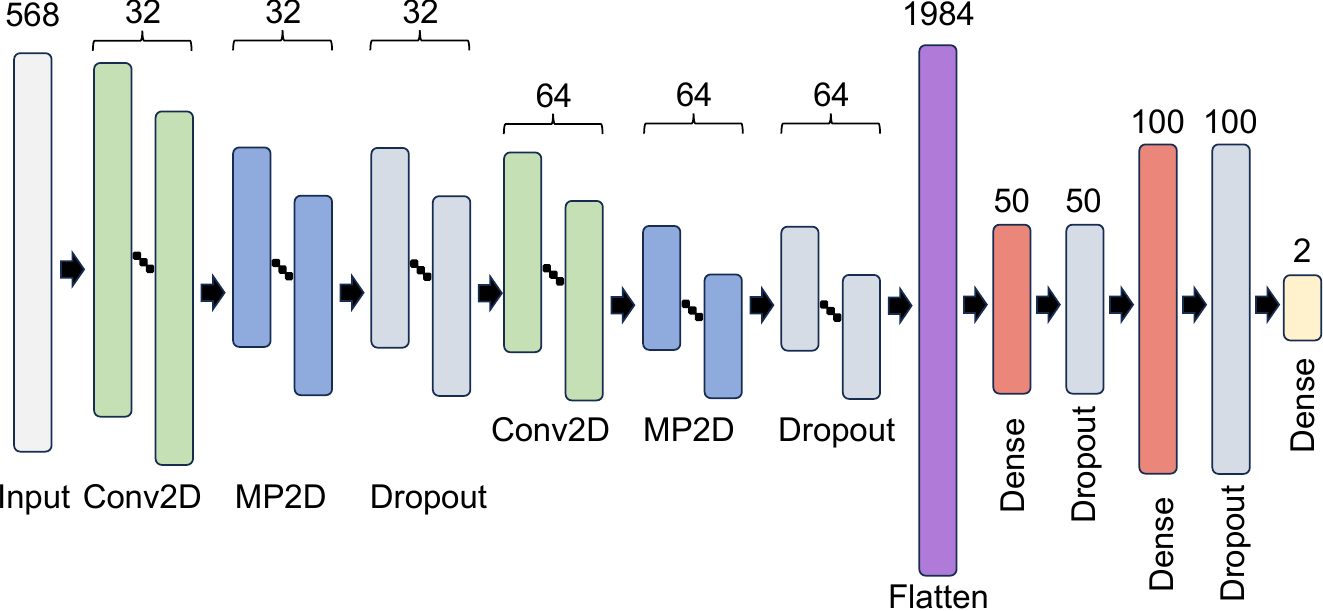}
    
    \caption{The DNN architecture used in \namex{} with two Conv1D layers (Conv2D with the second dimension as one), followed by a flatten and three dense layers.} \label{fig:neuralnetwork}
\end{figure}

\begin{figure}[!t]
    \centering
    \vspace*{0.03in}
    \includegraphics[width=0.45\textwidth]{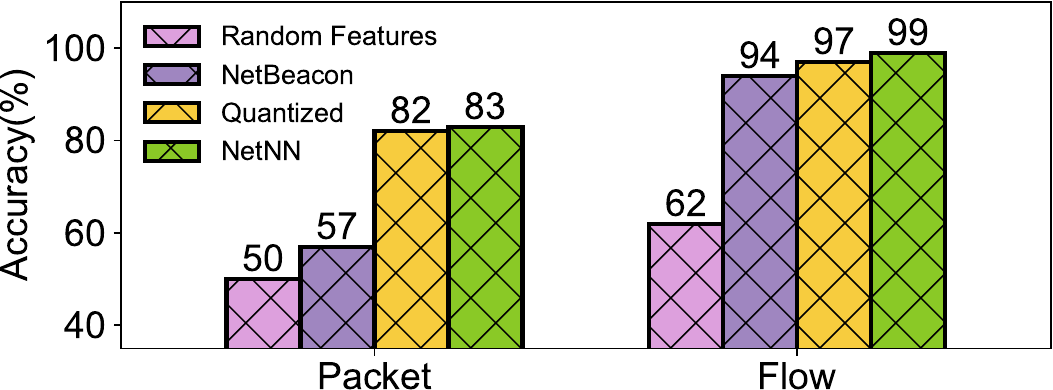}
    \caption{Comparison of NetBeacon, Random, and \namex for both packet and flow classification.} \label{fig:motivation_comparison_2}
\end{figure}

\noindent\textbf{Flow classification.} For the flow classification of \namex{}, we train a neural network (a 1D convolutional layer with 32 filters followed by another 1D convolutional layer with 64 filters, a dense layer with 50 neurons, another dense layer with 100 neurons, and a final output layer as depicted in Figure~\ref{fig:neuralnetwork}) with the first 68 bytes of the first power of two packets (1st, 2nd, 4th, ..., 1024th) with the minimum and maximum of inter-arrival times up to the 1024th packet. We injected a flow identifier as a feature to the input to help the neural network model generate similar outputs for the packets from the same flow. At the end, we aggregate the classification results of the ten packets and report the final category of the flow based on the majority votes. For NetBeacon, similar to FlowLens~\cite{barradas2021flowlens}, we use buckets for length and inter-arrival time distributions of all the packets in the dataset, roughly 14 million packets (7 million malicious and 7 million benign). Specifically, we use 1500 buckets for different lengths (1, 2, ..., 1500) and 3000 buckets for inter-arrival times (0-60 seconds distributed to 3000 buckets with lengths of 12 milliseconds). We randomly select $10\%$ of lengths and inter-arrival times buckets for the random feature approach. With the flow level features, Netbeacon's accuracy reaches $94\%$ while selecting a random subset of features reduces the accuracy considerably to $62\%$. On the contrary, \namex{} achieves a validation accuracy of over $99\%$ with less information as shown in Figure~\ref{fig:motivation_comparison_2}.

Flow-level feature extraction is quite expensive in the data plane as accurately calculating statistical features of packets such as inter-arrival rate or length distribution requires floating-point arithmetic, which is not available on the programmable switches data plane such as Intel Tofino. Moreover, storing and updating features require careful design and management of multiple hash tables~\cite{barradas2021flowlens, zhou2023efficient, akem2023flowrest}. On the other hand, we observe the importance of careful feature selection by using a random set of 50 features from FlowLens~\cite{barradas2021flowlens}, resulting in the lowest accuracy. The above combined raises the question of how to reduce the overhead of feature extraction.

\subsection{Our Goal}
Our goal is to design a new intrusion detection system based on DNNs that run completely in the network data plane, leveraging the performance of modern programmable switches. With that, a network operator can scan network traffic at the line rate and analyze it accurately directly in the network data plane. The following objectives encapsulate the underlying motivations guiding our design principles.

\noindent\textbf{No feature engineering.} We aim to avoid the use of features for packet/flow classification since gathering features is both computing and storage expensive, as we discuss in Section~\ref{sec:feature_extraction}. Therefore, our objective is mainly to use raw packet representation plus inter-arrival time information to feed a DNN model for intrusion detection.

\noindent\textbf{Model scalability.} We aim to deploy large DNNs with hundreds of thousands of operations per inference for high accuracy, which is not possible on only one switch as discussed in Section~\ref{sec:switch_limitation}. Therefore, we propose to deploy a DNN by breaking it into multiple parts and distributing each part's weights to a different programmable switch similar to~\cite{razavi2022distributed}.

\subsection{Intrusion Detection on Programmable Switches} \label{sec:switch_limitation}
Intrusion detection on programmable switches (e.g., based on decision trees) has become popular due to the increasing demand for real-time network security~\cite{barradas2021flowlens, zhou2023efficient, zheng2021planter, zheng2022iisy, liu2021jaqen}. 
By performing intrusion detection directly in the data plane, the processing overhead and the network traffic can be significantly reduced, resulting in faster and more efficient intrusion detection. Additionally, since the programmable switches are located at strategic points in the network topology, they can provide better visibility into the network traffic, making it easier to detect and prevent network attacks. 

\noindent\textbf{Processing constraint.} To ensure line-rate processing in programmable switches, the data plane programs (e.g., written in P4) must employ simple packet processing instructions. Each pipeline stage in the switch imposes a fixed processing time on packets. This restriction limits the number and complexity of operations performed within each stage. Floating-point operations, loops, and even integer multiplication/division are usually not supported. Also, the actions associated with each table are constrained to restricted simpler operations, such as additions and bit shifts.

\noindent\textbf{Memory constraint.} The memory architecture of programmable switches imposes various constraints on the data structures used in P4 programs. Such switches incorporate two high-speed memory types: TCAM (Ternary Content Addressable Memory), which facilitates rapid table lookups, and SRAM (Static Random Access Memory), which allows P4 programs to maintain state across packets. However, the available stateful memory in programmable switches is limited, typically in 10s of MB. Also, accessing all available registers can be challenging since registers within one stage of the processing pipeline cannot be accessed from different stages.

\section{System Design} \label{sec:design}
\begin{figure}[!t]
    \centering
    \vspace*{0.03in}
    \includegraphics[width=0.45\textwidth]{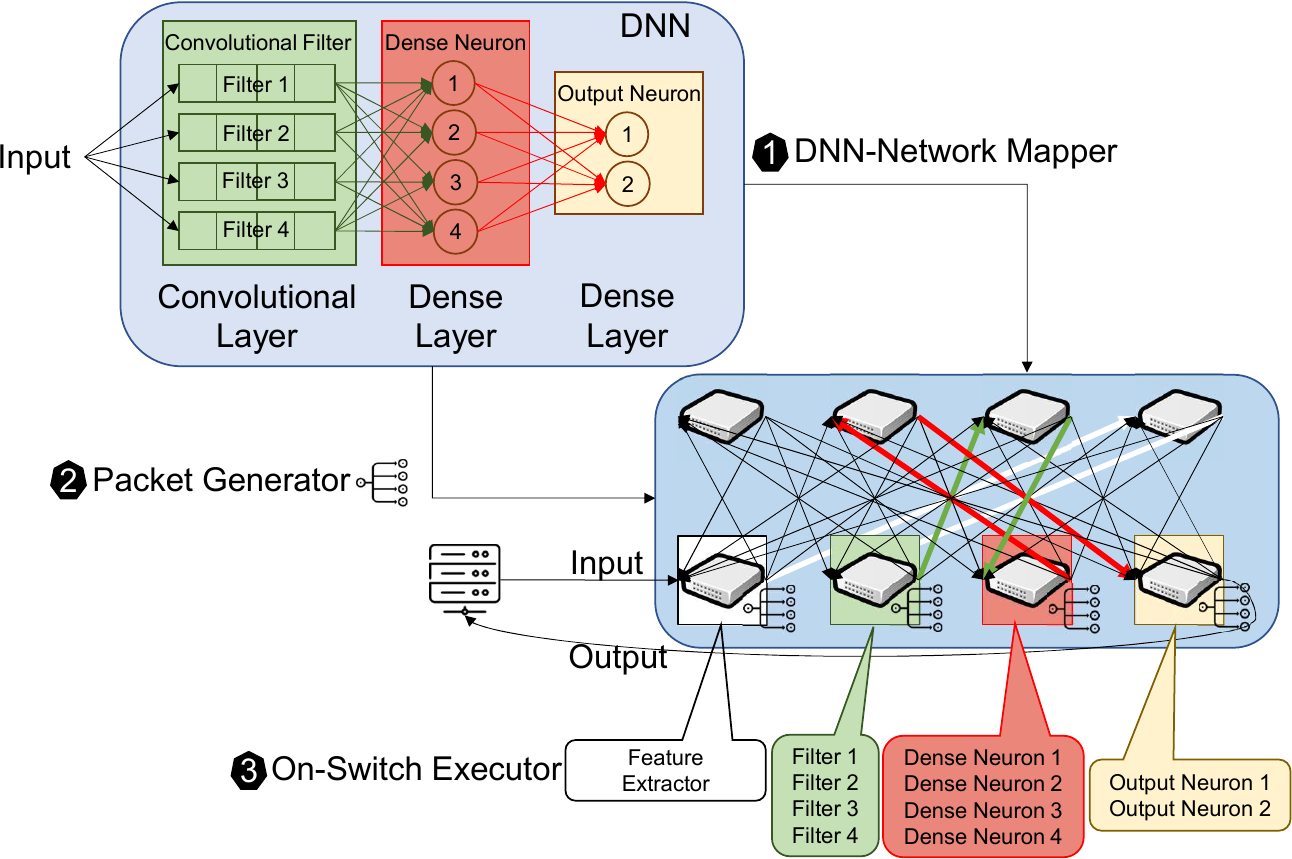}
    
    \caption{An overview of the \namex{} architecture.} \label{fig:system_design}
\end{figure}

We present \namex{}, a system that leverages the network data plane for fast and accurate intrusion detection. We start with a design overview and then dive into the major components. 

\subsection{System Overview}

A DNN is characterized by its layered architecture, consisting of multiple interconnected layers of neurons. This architecture typically includes an initial layer, one or more intermediate layers, and a final output layer. Within these layers, neurons are connected via weighted connections, allowing for complex information processing. 
This work focuses on three different layer types, Convolutional, MaxPooling, and Dense layers, where each of the layers has different properties such as convolutional kernel and maxpooling window.

Figure~\ref{fig:system_design} (top left) shows a simple DNN and provides a high-level overview of \namex{} consisting of three key components: \textit{Mapper}, \textit{Packet Generator}, and \textit{Neural Network Executor}. 
The system takes a trained model as input and decides the number of needed switches by leveraging the layered architecture of the DNNs. 
Take the model in Figure~\ref{fig:neuralnetwork} as an example. After the model has been provided to the system, the system decides the number of switches, in this case, layer by layer (Convolutional and Dense layers) per switch. Moreover, since switches have multiple pipelines to process packets in parallel
, we divide the convolutional layers filter by filter and, divide the dense neurons by the number of neurons and map each partition to a switch pipeline (details in Section~\ref{sec:mapper}). 

When a packet enters the \textit{Feature Extractor} switch 
we first extract the packet identifier (5-tuple: \texttt{src.ip}, \texttt{src.port}, \texttt{dst.ip}, \texttt{dst.port}, and \texttt{proto}) and capture the packet's arrival time. Then, we check if the packet belongs to a flow recorded before. If it is recorded in the flow table, we calculate the simple inter-arrival time features of the flow up to the current packet. We further check whether the current packet should be treated as an inference point (see Section~\ref{sec:executor}). If so, we extract the packet's first $68$ bytes (maximum UDP/IP datagram size) and take each bit as input to our model, including the inter-arrival time features. 
When the features are preprocessed, the feature extractor switch generates packets from the set of features based on the next layer type with respect to the network topology and the mappings as described in Section~\ref{sec:generator}. The packets are generated using the convolution kernel if the next layer is a convolutional layer. 
Finally, we execute the DNN operations based on the layer type (see Section~\ref{sec:executor}) using simple arithmetic operations (bit-shifting and addition). If the layer is convolutional, we dot product the neurons by the link weights and aggregate the results. After all the results are computed for a layer, a similar approach for the first layer packet generation is applied to the rest of the layers, e.g., the results are treated as new packets, and based on the next layer type, the packets are generated and routed.

\subsection{Mapper} \label{sec:mapper}
In this part, we focus on the DNN partition problem to fit the model on a set of programmable switches and provide an efficient execution flow of the DNN. Most DNNs have a layered structure, and we decide to partition the DNNs across switches layer by layer, as done similarly in~\cite{razavi2022distributed}. We follow the same practice and assign each layer to a switch. There are layers without weights, such as MaxPooling layers, typically employed after convolutional layers to reduce overfitting and increase computational efficiency. MaxPooling layers, unlike other layers in a neural network, do not involve learnable weight parameters. Instead, they perform computations based on a fixed operation to downsample and retain the most important information from the input data. Since these layers do not have weight parameters, we fit them alongside their previous convolutional layers. After mapping the layers/neurons to switches, we store the model weights on switches as follows. 


\begin{figure*}[!t]
    \centering
    \vspace*{0.03in}
    \includegraphics[width=0.95\textwidth]{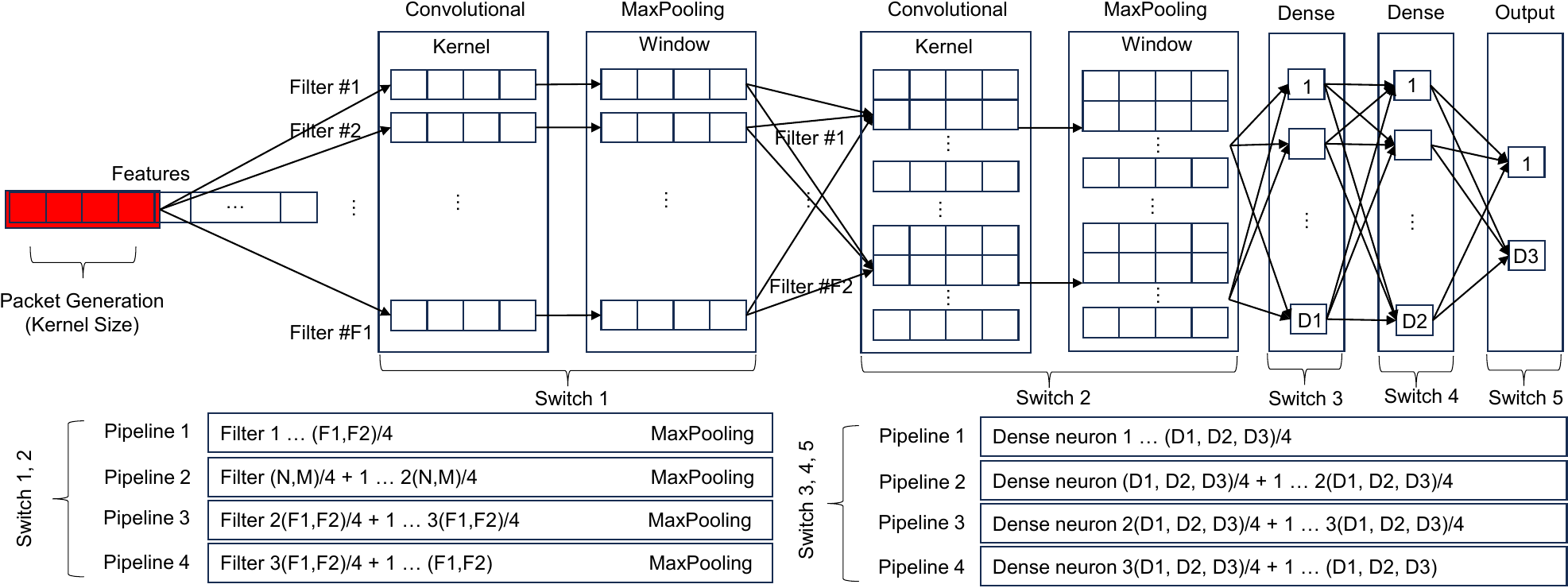}
    
    \caption{The workings of \namex{} Mapper.} \label{fig:mapper_design}
\end{figure*}

\noindent\textbf{Convolutional layers.} We divide the filters by the pipelines on a switch. The filter-splitting approach suits the switch architecture well since there is no communication between any two filters in the same convolutional layer, and computations can be done in parallel. For the model in Figure~\ref{fig:neuralnetwork}, each pipeline in the first switch receives two filters, while the second switch receives four filters per pipeline. In the inference phase, after the convolutional layers, there are MaxPooling layers, which will fit in the same switches as shown in Figure~\ref{fig:mapper_design}. 

There are two different approaches to storing the weights of convolution layers: register arrays in the data plane and table entries through the control plane. With register arrays, we can read and write them in real-time during packet processing since it does not involve software processing. However, register arrays often have limited capacity due to hardware constraints, and they are typically designed for transient data storage needed during packet processing and not for persistent data. With table entries, we benefit from simplified management of them since the modification is usually handled through APIs. However, accessing and processing entries in tables via the control plane can introduce a higher latency compared to data plane register arrays. This latency is due to software processing and lookup operations, making tables less suitable for real-time packet processing. We use table entries to access the persistent data (layers' weights) and register arrays for storing data(features)/intermediate data in real time.

To store the weights of the convolutional layers, we use the filter number and the position of the weight in that filter ($Filter, Weigth_i, Weight_j$) as the identifier. However, we need a more complex 3D table entry on the switch. Therefore, we need to map the 3D identifier to a 1D space, which is usually done by their multiplication, $Filter \times Height \times Width$, that maps a number from $\mathbb{R}^3 \longrightarrow \mathbb{R}$. However, calculating the multiplications on the data plane is computationally expensive as multiplications are not natively supported on the programmable switches. Therefore, we use an available hash function on the data plane to create a unique index using similar identifiers, i.e., Hash$(Filter \times Height \times Width)$. 

\noindent\textbf{Dense layers.} We follow the same approach as in the convolutional layer, 
we divide the layer by the number of dense neurons. Similar arguments apply here as there is no communication between two dense neurons in the same layer, and the dense neuron computations can be executed in parallel. 
We use a similar data structure (table entry) for storing the weights of dense layers. However, storing dense weights is simpler than convolutional weights as one can access the dense weights by a number (weight identifier) instead of three numbers.

So far, we have mapped a DNN with different layers
to a set of switches. Next, we need to trigger the DNN computations through the events of packet arrivals. When a packet arrives at a switch, we need to mimic the DNN execution using network packets. We first identify the next layer where the input/intermediate data should be sent. Based on the layer type, we follow different strategies:

\noindent\textbf{Convolutional layers.} If the next layer type is a convolutional layer, we put the related data based on the kernel size of the convolutional layer to the packet header (including the necessary identifiers) and multicast the packet to all the next switch pipelines. For instance, from the feature extractor switch, we fetch a kernel size of features from the feature register, add the filter number and weight positions in the kernel ($kernel_i, kernel_j$) to the packet header, and create a packet. The result of this approach is that when a packet arrives at a convolutional switch, it can perform all the necessary computations related to that exact packet header data, except the MaxPooling part (discussed in Section~\ref{sec:executor}) as it requires multiple packet headers. When the data in the packet headers is processed, we get a final value for the current packet header that will be used to generate the next packet header based on the following layer type.

\noindent\textbf{Dense layers.} If the next layer is a dense layer, we multicast each feature to all the next switch pipelines since each feature is used by all the neurons of a dense layer. After all the computations regarding the current data are completed, the result 
is stored in a register array on the switch. After all the computations regarding a dense neuron are finished, the finalized value is ready to be sent to the pipelines of the next switch based on the next layer type.

\subsection{Neural Network Executor} \label{sec:executor}
When a packet arrives in the initial switch, we capture the inter-arrival time and calculate the simple inter-arrival time features (min, max) up to the current packet. After that, we need to identify if the current packet is an inference point. Following NetBeacon~\cite{zhou2023efficient}, we check whether a packet is in the order of the power of two, namely (1, 2, 4, ...), as it can be easily identified in a programmable switch by using a bitwise \texttt{AND} operation between the number and its decrement (if the result is zero, it means that the number is a power of two). Upon an inference point, we start generating packets as described before. We use the following procedures (based on the layer type) to process the generated packets.

\noindent\textbf{Convolutional and MaxPooling layers.} For convolutional layers, we first create the hash of the indices (filter number, height, and width) of the required weights. After fetching the weights from registers, we perform dot products for all the weights and the values. We break down each multiplication operation into a series of bit shifts for the dot production. Specifically, we shift the input left by an amount determined by the corresponding bit in the 8-bit weight. These shifts are executed concurrently in a single processing stage, and the interim results are stored temporarily. Then, a reduction step combines the interim results. For 8-bit weights, this step necessitates three stages. Performing integer multiplication in the data plane is viable for executing an inference, but it takes too many switch resources (stages). Therefore, we adopt an alternative approach where we store all the dot product results in a table to reduce the number of required stages. We store all possible dot products with 8-bit to 8-bit values and use a TCAM table to access the dot product with value|weight. This approach requires one stage instead of three stages for the dot product with the memory requirement of $2^{16}$ 8-bit values (65~KB of memory). Next, we employ the ReLU activation function to process the dot product result, which involves examining the most significant bit (MSB), 
the sign bit, and substituting the value with 0 when it is set to 1.

To perform MaxPooling operations, we need to be sure that all the values regarding a MaxPooling window (a MaxPooling window is a rectangular region of the values where the most important value is selected) have arrived at the switch. To ensure that, we use a bitmap (using a table) of each value and flag its indexes when they arrive. Anytime a packet arrives, we check the bitmap of the value. If all packets have arrived, we fetch the results of all the packets from the convolutional layer regarding the same MaxPooling window and find the maximum of the values within $log(size(MaxPooling window))$. After calculating the maximum value in the MaxPooling window
, we multicast it to the following switch pipelines.

\noindent\textbf{Dense layers.} For the dense layer, finding the index of the dense neuron is relatively straightforward. When a packet arrives at a switch running part of a dense layer, we use a counter to access the dense table entry (the data structure we use for storing the dense weights) to iterate over all the dense weights. After the calculation, we store the intermediate results in a register array and aggregate them as soon as the processing associated with that neuron is finished. We repeat the procedure until all variables from the previous layer are processed.
After that, we perform a ReLU activation function (explained above) to see if we want to activate the neuron. Finally, we multicast the result to the next switch pipelines.

\subsection{Packet Generator} \label{sec:generator}

\begin{figure}[!t]
    \begin{subfigure}{0.45\textwidth}
      \centering
      \includegraphics[width=1\linewidth]{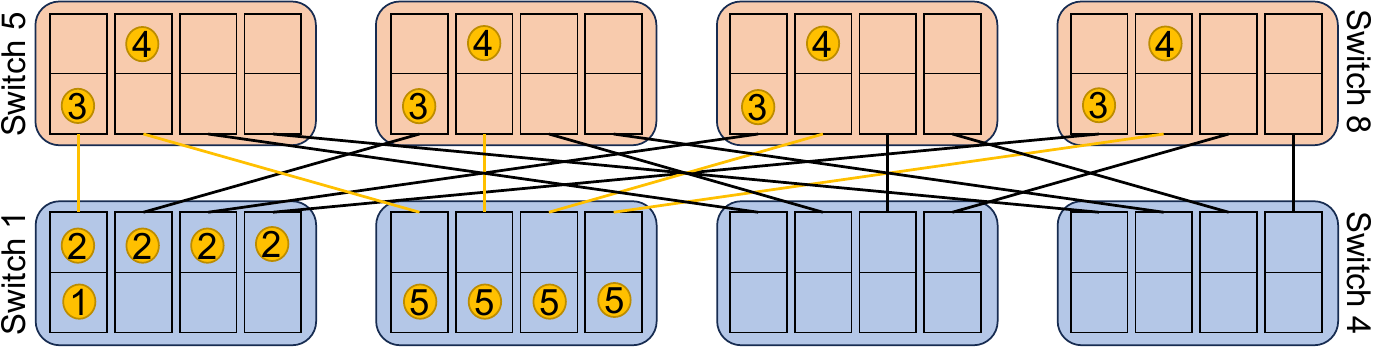}
      \caption{2-tier Clos topology}
      \label{fig:2-tier}
    \end{subfigure}
    \hfill
    \begin{subfigure}{0.45\textwidth}
      \centering
      \includegraphics[width=1\linewidth]{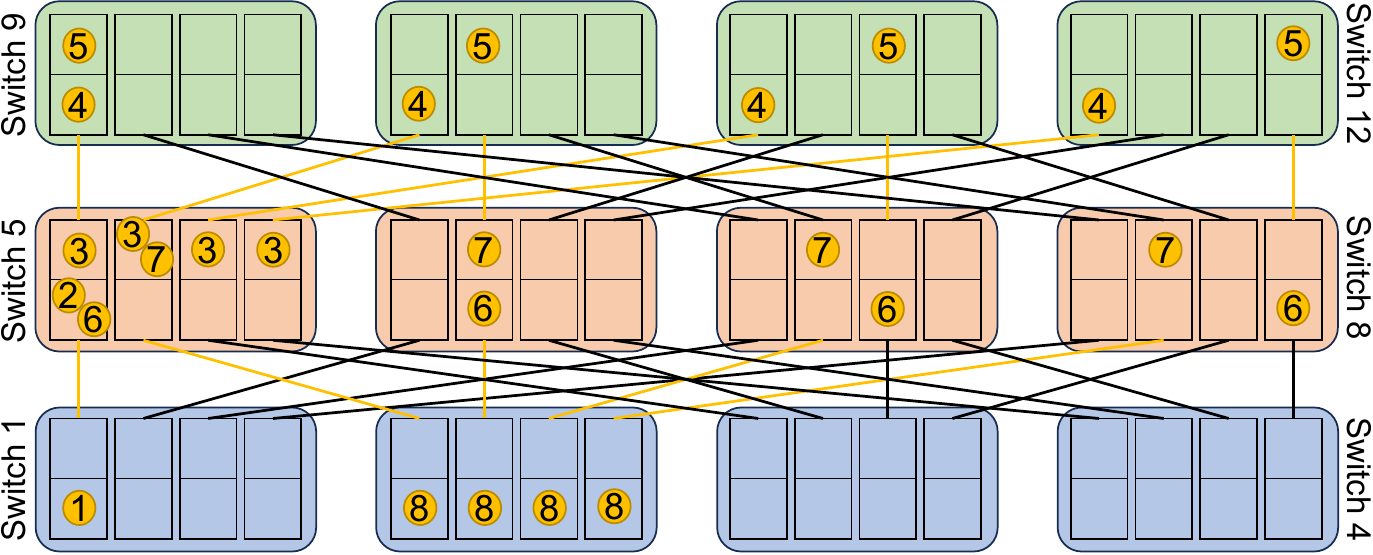}
      \caption{3-tier Clos topology}
      \label{fig:3-tier}
    \end{subfigure}
    \caption{The target network architecture. 
    } \label{fig:network_design}
\end{figure}

The data center network architecture we use is depicted in Figure~\ref{fig:network_design}, which represents a multi-tier Clos network topology~\cite{singh2015jupiter}. Each tier has four programmable switches, each with four pipelines with ingress and egress parts.
Each switch in a tier is connected to all the switches in the above and bellow tiers (if available) with 400~GbE links. For instance, there is exactly one link from \textit{Switch 1} (pipeline 4) in the first tier to \textit{Switch 8} (pipeline 1) in the second tier. The advantage of Clos is that every switch from each tier is exactly two hops away from another switch in the same tier. Leveraging this property, we map the layers of the DNN exclusively to the lowest tier switches, mitigating synchronization issues and eliminating potential stragglers within the network. The downside of this architecture is that there is no direct link between all the pipelines of a switch in a tier to another switch in the above or below tier. Therefore, we may need another tier to send packets to the correct switch pipeline. In a similar example, there is no direct link between pipeline 1 of \textit{Switch 1} to pipeline 1 of \textit{Switch 8}. Consequently, if we want to have the packet emitting from pipeline 1 of \textit{Switch 1} in ingress part of pipeline 1 of \textit{Switch 8}, we need to send the packet to a middle switch and then get the packet from the middle switch that is directly connected to pipeline 1 of \textit{Switch 8}, which in this scenario are pipeline 4 of \textit{Switch 1} and pipeline 4 of \textit{Switch 9}.

After a packet arrives in a switch (say \textit{Switch 1} (\yellowballnumber{1}), the processing part will start in the ingress part of the switch. 
If all the processing related to the packet is finished in the ingress (see Section~\ref{sec:executor}), we use a 2-tier Clos network architecture as illustrated in Figure~\ref{fig:2-tier}.
The result in the ingress part will be distributed to all egress parts of the switch using the traffic manager (\yellowballnumber{2}). Next, 
the result is encapsulated into packets in the egress part and the packets are emitted to the switches in a higher tier (\yellowballnumber{3}). Then, the switches in the higher tier direct the packets to the corresponding pipelines to be sent to a specific pipeline in the lower tier switch (\yellowballnumber{4}). Finally, all the pipelines of \textit{Switch 2} receive the packets with the result from the initial switch (\textit{Switch 1})) (\yellowballnumber{5}). All the associated links are colored. 

However, if the processing part requires both the ingress and ingress part of the pipeline, we may need another tier to multicast the result, as illustrated in Figure~\ref{fig:3-tier}. Taking a similar scenario, sending the result of a packet in pipeline 1 of \textit{Switch 1} to all pipelines of \textit{Switch 2}, instead of distributing the packet in \textit{Switch 1}, we need to send the packet encapsulating the result to the second tier switch (\yellowballnumber{2}), \textit{Switch 5} here, and distribute it to all pipelines using the traffic manager in the switch (\yellowballnumber{3}). After that, we follow the similar steps as a 2-tier Clos architecture until all the packets are in the correct pipelines of the second-tier switches (\yellowballnumber{4}, \yellowballnumber{5}, \yellowballnumber{6}, \yellowballnumber{7}). One of the second-tier switches has the packet in the correct pipeline without sending it to a higher-tier switch (pipeline 2 of \textit{Switch 5}). To avoid synchronization issues, we follow the same procedure for all the second-tier switches. Finally, we send the packets to all pipelines of \textit{Switch 2} (\yellowballnumber{8}). All the used links are colored in this scenario as well.

\section{Evaluation} \label{sec:evaluation}
We implemented a preliminary version of \namex{} using P4 language with the behavioral model bmv2, which serves as a P4 target. This prototype is used for testing within Mininet, a network emulator tailored for network experimentation. Within Mininet, we established a network topology illustrated in Figure~\ref{fig:system_design} with four switches per layer (since there are four pipelines per switch in a Tofino switch) as depicted in Figure~\ref{fig:mapper_design}, and switches are connected to each other similar to the link connections in Figure~\ref{fig:3-tier}.

\subsection{Resource Usage}
By executing the P4 implementation on hardware, we can achieve the switch's line-rate performance and naturally leverage the match-action pipeline's characteristics, including bounded latency guarantees and high-speed throughput. However, in our prototype, the software emulation on bmv2 lacks the necessary time accuracy for precise performance measurements at a granular level. Therefore, we report the memory usage, the number of operations, and the packet generation rates of the \namex{} implementation. As presented in Table~\ref{table:performance}, the convolutional layers require much more operations per packet than dense layers, hence more switches are beneficial to distribute the computations. On the other hand, dense layers require more memories to store the weights, yet much lower than a programmable switch memory limitation.

\subsection{Impact of Parameters}
    

    

\begin{figure}
	\centering
    \vspace*{0.03in}
	\begin{minipage}{.49\columnwidth}
		\centering
		\includegraphics[width=\textwidth]{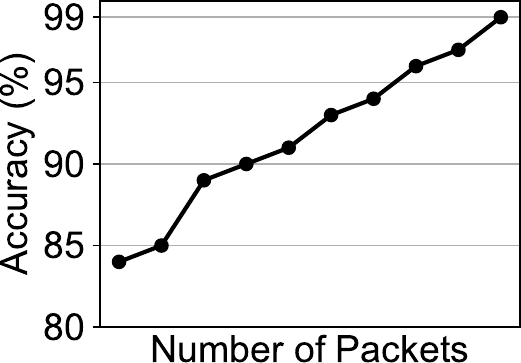}
		\caption{Inference points.}
		\label{fig:inference_point}
	\end{minipage}%
	\begin{minipage}{.49\columnwidth}
		\centering
		\includegraphics[width=\textwidth]{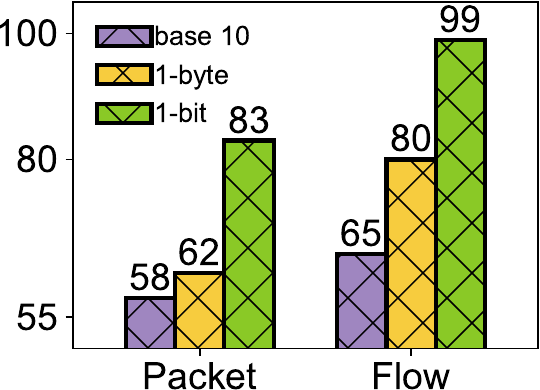}
		\caption{Input representation.}
		\label{fig:bit_byte}
	\end{minipage}
\end{figure}

\begin{table}
    \centering
    \footnotesize
    \caption{DNN in-switch performance. Memory is in byte.}
    \small
    \begin{tabular}{lrrr}
         \toprule
          \textbf{Layer} & \textbf{$\#$ of packets} & \textbf{Operations per packet} & \textbf{Memory} \\
         \midrule
         Conv1  & 3240 & 644   & 1296  \\
         Conv2  & 2592 & 38895 & 5680  \\
         Dense1 & 496  & 3159  & 25792 \\
         Dense2 & 4    & 4950  & 1250  \\
         Dense3 & 1    & 378   & 51    \\
         \bottomrule
    \end{tabular}
    \label{table:performance}
\end{table}

\noindent\textbf{Impact of inference points.} The number of inference points used to determine classification is an important factor affecting performance. The more packets from the same flow that we observe and analyze, the more accurate the features at the flow level become, and hence, the more accurate the flow classification confidence. As depicted in Figure~\ref{fig:inference_point}, a clear trend shows that the accuracy improves as we incorporate more inference points for the flow classification. If the goal of the classification is 90\% accuracy, the system needs at least four packets of the same flow; if the goal is to reach over 95\%, the system requires at least eight packets to be analyzed.

\noindent\textbf{Impact of input representation.} 
To fine-tune the DNN for intrusion detection, we used different input formats to improve the model's accuracy. We trained the same model with three different input formats: bit, byte, and base 10 to demonstrate the model accuracy with different data representations. Figure~\ref{fig:bit_byte} illustrates the accuracy of the model trained with different input formats. Since the number of flows in the dataset is limited (1000 flows) and each flow provides few features, the model trained with base 10 format not only provides the least accuracy 
 in both packet and flow classification but also requires feature engineering (calculating the features in base 10), while the model trained with 1-bit format provides the highest accuracy without the need for feature engineering since that is the natural format of network packets.
\section{Related Work} \label{sec:related_work}
\noindent\textbf{Intrusion detection.}
Due to switches' strategic location and performance, there is an increased usage of switches for intrusion detection~\cite{xing2020netwarden, fu2021realtime, mirsky2018kitsune}.
FlowLens~\cite{barradas2021flowlens} leverages programmable switches to support machine learning-based network security applications by collecting features of packet distributions and classifying flows on the switches. 
Jaqen~\cite{liu2021jaqen} addresses the challenge of defending against volumetric DDoS attacks by leveraging universal sketch techniques, switch-native mitigation, and network-wide management.
Poseidon~\cite{zhang2020poseidon} provides a modular policy abstraction for DDoS defense policies.
NetBeacon~\cite{zhou2023efficient} proposes a multi-phase sequential model architecture for dynamic analysis using a switch-based model representation.

\noindent\textbf{ML-based applications.} In-network ML has gained significant attentions~\cite{2021-nsdi-switchml, 2019-hotnets-classification}.
IIsy~\cite{zheng2022iisy} is a framework for in-network classification using off-the-shelf network devices. It maps trained ML models to network devices without modifications.
FPISA~\cite{yuan2022unlocking} proposes a floating point representation in programmable switches for efficient distributed training.
Mousika~\cite{xie2022mousika} proposes a Binary Decision Tree (BDT) model that enables translation from complex models to BDT using knowledge distillation.
N3IC~\cite{2022-nsdi-n3ic} enables neural network inference on programmable NICs.
Planter~\cite{zheng2021planter} maps ensemble models to programmable switches for efficient data classification.
SwitchTree~\cite{lee2020switchtree}, pForest~\cite{busse2019pforest}, and FlowRest~\cite{akem2023flowrest} enhance in-switch inference with Random Forest models.

None of these works directly apply DNNs for intrusion detection completely in the network data plane.

\section{Conclusion} \label{sec:conclusion}
In this paper, we presented \namex{}, a novel approach to enable the execution of DNNs in programmable switches. \namex{} achieves this by splitting and distributing the neural network layers to multiple switches, generating packets similar to the flow execution of the neural network, and simplifying the needed mathematical operations for getting the inference based on the capabilities of the programmable switches. We prototype \namex{} using P4 and evaluate it by designing a no-feature-engineering-needed DNN using a Covert Channel dataset, showing an intrusion detection accuracy of 99\%.
\section*{Acknowledgment}\label{sec:acknowledment}
We would like to thank the anonymous reviewers for their valuable comments and suggestions. This work has been funded by Deutsche Forschungsgemeinschaft (DFG, German Research Foundation) – Project-ID 210487104 - SFB 1053.

\bibliographystyle{ieeetran}
\bibliography{ref}

\begin{thebibliography}{10}
\providecommand{\url}[1]{#1}
\csname url@samestyle\endcsname
\providecommand{\newblock}{\relax}
\providecommand{\bibinfo}[2]{#2}
\providecommand{\BIBentrySTDinterwordspacing}{\spaceskip=0pt\relax}
\providecommand{\BIBentryALTinterwordstretchfactor}{4}
\providecommand{\BIBentryALTinterwordspacing}{\spaceskip=\fontdimen2\font plus
\BIBentryALTinterwordstretchfactor\fontdimen3\font minus
  \fontdimen4\font\relax}
\providecommand{\BIBforeignlanguage}[2]{{%
\expandafter\ifx\csname l@#1\endcsname\relax
\typeout{** WARNING: IEEEtran.bst: No hyphenation pattern has been}%
\typeout{** loaded for the language `#1'. Using the pattern for}%
\typeout{** the default language instead.}%
\else
\language=\csname l@#1\endcsname
\fi
#2}}
\providecommand{\BIBdecl}{\relax}
\BIBdecl

\bibitem{zhang2020poseidon}
M.~Zhang, G.~Li, S.~Wang, C.~Liu, A.~Chen, H.~Hu, G.~Gu, Q.~Li, M.~Xu, and
  J.~Wu, ``Poseidon: Mitigating volumetric ddos attacks with programmable
  switches,'' in \emph{the 27th Network and Distributed System Security
  Symposium (NDSS 2020)}, 2020.

\bibitem{liu2021jaqen}
Z.~Liu, H.~Namkung, G.~Nikolaidis, J.~Lee, C.~Kim, X.~Jin, V.~Braverman, M.~Yu,
  and V.~Sekar, ``Jaqen: A high-performance switch-native approach for
  detecting and mitigating volumetric ddos attacks with programmable
  switches,'' in \emph{30th USENIX Security Symposium (USENIX Security 21)},
  2021, pp. 3829--3846.

\bibitem{barradas2021flowlens}
D.~Barradas, N.~Santos, L.~Rodrigues, S.~Signorello, F.~M. Ramos, and
  A.~Madeira, ``Flowlens: Enabling efficient flow classification for ml-based
  network security applications.'' in \emph{NDSS}, 2021.

\bibitem{zheng2022iisy}
C.~Zheng, Z.~Xiong, T.~T. Bui, S.~Kaupmees, R.~Bensoussane, A.~Bernabeu,
  S.~Vargaftik, Y.~Ben-Itzhak, and N.~Zilberman, ``Iisy: Practical in-network
  classification,'' \emph{arXiv preprint arXiv:2205.08243}, 2022.

\bibitem{zhou2023efficient}
G.~Zhou, Z.~Liu, C.~Fu, Q.~Li, and K.~Xu, ``An efficient design of intelligent
  network data plane,'' in \emph{32nd USENIX Security Symposium (USENIX
  Security 23)}, 2023, pp. 6203--6220.

\bibitem{akem2023flowrest}
A.~T.-J. Akem, M.~Gucciardo, and M.~Fiore, ``Flowrest: Practical flow-level
  inference in programmable switches with random forests,'' in \emph{IEEE
  INFOCOM 2023-IEEE Conference on Computer Communications}.\hskip 1em plus
  0.5em minus 0.4em\relax IEEE, 2023, pp. 1--10.

\bibitem{2020-osdi-serving}
A.~Gujarati, R.~Karimi, S.~Alzayat, W.~Hao, A.~Kaufmann, Y.~Vigfusson, and
  J.~Mace, ``Serving dnns like clockwork: Performance predictability from the
  bottom up,'' in \emph{14th USENIX Symposium on Operating Systems Design and
  Implementation (OSDI 20)}, 2020, pp. 443--462.

\bibitem{razavi2022fa2}
K.~Razavi, M.~Luthra, B.~Koldehofe, M.~M{\"u}hlh{\"a}user, and L.~Wang, ``Fa2:
  Fast, accurate autoscaling for serving deep learning inference with sla
  guarantees,'' in \emph{2022 IEEE 28th Real-Time and Embedded Technology and
  Applications Symposium (RTAS)}.\hskip 1em plus 0.5em minus 0.4em\relax IEEE,
  2022, pp. 146--159.

\bibitem{xing2020netwarden}
J.~Xing, Q.~Kang, and A.~Chen, ``Netwarden: Mitigating network covert channels
  while preserving performance,'' in \emph{29th USENIX Security Symposium
  (USENIX Security 20)}, 2020, pp. 2039--2056.

\bibitem{doriguzzi2024introducing}
R.~Doriguzzi-Corin, L.~A.~D. Knob, L.~Mendozzi, D.~Siracusa, and M.~Savi,
  ``Introducing packet-level analysis in programmable data planes to advance
  network intrusion detection,'' \emph{Computer Networks}, vol. 239, p. 110162,
  2024.

\bibitem{2022-asplos-taurus}
\BIBentryALTinterwordspacing
T.~Swamy, A.~Rucker, M.~Shahbaz, I.~Gaur, and K.~Olukotun, ``Taurus: a data
  plane architecture for per-packet {ML},'' in \emph{{ASPLOS} '22: 27th {ACM}
  International Conference on Architectural Support for Programming Languages
  and Operating Systems, Lausanne, Switzerland, 28 February 2022 - 4 March
  2022}, B.~Falsafi, M.~Ferdman, S.~Lu, and T.~F. Wenisch, Eds.\hskip 1em plus
  0.5em minus 0.4em\relax {ACM}, 2022, pp. 1099--1114. [Online]. Available:
  \url{https://doi.org/10.1145/3503222.3507726}
\BIBentrySTDinterwordspacing

\bibitem{2019-hotnets-classification}
\BIBentryALTinterwordspacing
Z.~Xiong and N.~Zilberman, ``Do switches dream of machine learning?: Toward
  in-network classification,'' in \emph{Proceedings of the 18th {ACM} Workshop
  on Hot Topics in Networks, HotNets 2019, Princeton, NJ, USA, November 13-15,
  2019}.\hskip 1em plus 0.5em minus 0.4em\relax {ACM}, 2019, pp. 25--33.
  [Online]. Available: \url{https://doi.org/10.1145/3365609.3365864}
\BIBentrySTDinterwordspacing

\bibitem{2022-nsdi-n3ic}
G.~Siracusano, S.~Galea, D.~Sanvito, M.~Malekzadeh, G.~Antichi, P.~Costa,
  H.~Haddadi, and R.~Bifulco, ``Re-architecting traffic analysis with neural
  network interface cards,'' in \emph{19th {USENIX} Symposium on Networked
  Systems Design and Implementation, {NSDI} 2022, April 4-6, 2022}.\hskip 1em
  plus 0.5em minus 0.4em\relax {USENIX} Association, 2022, pp. 513--533.

\bibitem{razavi2022distributed}
K.~Razavi, G.~Karlos, V.~Nigade, M.~M{\"u}hlh{\"a}user, and L.~Wang,
  ``Distributed dnn serving in the network data plane,'' in \emph{Proceedings
  of the 5th International Workshop on P4 in Europe}, 2022, pp. 67--70.

\bibitem{2017-sosp-netcache}
\BIBentryALTinterwordspacing
X.~Jin, X.~Li, H.~Zhang, R.~Soul{\'{e}}, J.~Lee, N.~Foster, C.~Kim, and
  I.~Stoica, ``Netcache: Balancing key-value stores with fast in-network
  caching,'' in \emph{Proceedings of the 26th Symposium on Operating Systems
  Principles, Shanghai, China, October 28-31, 2017}.\hskip 1em plus 0.5em minus
  0.4em\relax {ACM}, 2017, pp. 121--136. [Online]. Available:
  \url{https://doi.org/10.1145/3132747.3132764}
\BIBentrySTDinterwordspacing

\bibitem{2018-nsdi-netchain}
X.~Jin, X.~Li, H.~Zhang, N.~Foster, J.~Lee, R.~Soul{\'e}, C.~Kim, and
  I.~Stoica, ``Netchain:scale-freesub-rtt coordination,'' in \emph{15th USENIX
  Symposium on Networked Systems Design and Implementation (NSDI 18)}, 2018,
  pp. 35--49.

\bibitem{2021-nsdi-switchml}
A.~Sapio, M.~Canini, C.-Y. Ho, J.~Nelson, P.~Kalnis, C.~Kim, A.~Krishnamurthy,
  M.~Moshref, D.~Ports, and P.~Richt{\'a}rik, ``Scaling distributed machine
  learning with in-network aggregation,'' in \emph{18th USENIX Symposium on
  Networked Systems Design and Implementation (NSDI 21)}, 2021, pp. 785--808.

\bibitem{chen2016xgboost}
T.~Chen and C.~Guestrin, ``Xgboost: A scalable tree boosting system,'' in
  \emph{Proceedings of the 22nd acm sigkdd international conference on
  knowledge discovery and data mining}, 2016, pp. 785--794.

\bibitem{li2014facet}
S.~Li, M.~Schliep, and N.~Hopper, ``Facet: Streaming over videoconferencing for
  censorship circumvention,'' in \emph{Proceedings of the 13th Workshop on
  Privacy in the Electronic Society}, 2014, pp. 163--172.

\bibitem{barradas2017deltashaper}
D.~Barradas, N.~Santos, and L.~Rodrigues, ``Deltashaper: Enabling unobservable
  censorship-resistant tcp tunneling over videoconferencing streams,''
  \emph{Proceedings on Privacy Enhancing Technologies}, 2017.

\bibitem{marin2019deepsec}
G.~Mar{\'\i}n, P.~Casas, and G.~Capdehourat, ``Deepsec meets rawpower-deep
  learning for detection of network attacks using raw representations,''
  \emph{ACM SIGMETRICS Performance Evaluation Review}, vol.~46, no.~3, pp.
  147--150, 2019.

\bibitem{zheng2021planter}
C.~Zheng and N.~Zilberman, ``Planter: seeding trees within switches,'' in
  \emph{Proceedings of the SIGCOMM'21 Poster and Demo Sessions}, 2021, pp.
  12--14.

\bibitem{singh2015jupiter}
A.~Singh, J.~Ong, A.~Agarwal, G.~Anderson, A.~Armistead, R.~Bannon, S.~Boving,
  G.~Desai, B.~Felderman, P.~Germano \emph{et~al.}, ``Jupiter rising: A decade
  of clos topologies and centralized control in google's datacenter network,''
  \emph{ACM SIGCOMM computer communication review}, vol.~45, no.~4, pp.
  183--197, 2015.

\bibitem{fu2021realtime}
C.~Fu, Q.~Li, M.~Shen, and K.~Xu, ``Realtime robust malicious traffic detection
  via frequency domain analysis,'' in \emph{Proceedings of the 2021 ACM SIGSAC
  Conference on Computer and Communications Security}, 2021, pp. 3431--3446.

\bibitem{mirsky2018kitsune}
Y.~Mirsky, T.~Doitshman, Y.~Elovici, and A.~Shabtai, ``Kitsune: an ensemble of
  autoencoders for online network intrusion detection,'' \emph{arXiv preprint
  arXiv:1802.09089}, 2018.

\bibitem{yuan2022unlocking}
Y.~Yuan, O.~Alama, J.~Fei, J.~Nelson, D.~R. Ports, A.~Sapio, M.~Canini, and
  N.~S. Kim, ``Unlocking the power of inline floating-point operations on
  programmable switches,'' in \emph{19th USENIX Symposium on Networked Systems
  Design and Implementation (NSDI 22)}, 2022, pp. 683--700.

\bibitem{xie2022mousika}
G.~Xie, Q.~Li, Y.~Dong, G.~Duan, Y.~Jiang, and J.~Duan, ``Mousika: Enable
  general in-network intelligence in programmable switches by knowledge
  distillation,'' in \emph{IEEE INFOCOM 2022-IEEE Conference on Computer
  Communications}.\hskip 1em plus 0.5em minus 0.4em\relax IEEE, 2022, pp.
  1938--1947.

\bibitem{lee2020switchtree}
J.-H. Lee and K.~Singh, ``Switchtree: in-network computing and traffic analyses
  with random forests,'' \emph{Neural Computing and Applications}, pp. 1--12,
  2020.

\bibitem{busse2019pforest}
C.~Busse-Grawitz, R.~Meier, A.~Dietm{\"u}ller, T.~B{\"u}hler, and L.~Vanbever,
  ``pforest: In-network inference with random forests,'' \emph{arXiv preprint
  arXiv:1909.05680}, 2019.

\end{thebibliography}

\end{document}